\documentclass[%
 reprint,
superscriptaddress,%groupedaddress,%unsortedaddress,%runinaddress,%frontmatterverbose,%preprint,%showpacs,preprintnumbers,%nofootinbib,%nobibnotes,%bibnotes,
amsmath,amssymb,
]{revtex4-1}

\usepackage{graphicx}% Include figure files
\usepackage{dcolumn}% Align table columns on decimal point
\usepackage{bm}% bold math

\begin{document}

\preprint{APS/123-QED}

\title{Omni-resonant space-time wave packets}

\author{Abbas Shiri}
\author{Murat Yessenov}
\author{Rohinraj Aravindakshan}
\author{Ayman F. Abouraddy}

\email{Corresponding author: raddy@creol.ucf.edu}

\affiliation{CREOL, The College of Optics \& Photonics, University of Central Florida, Orlando, Florida 32186, USA}

% To be edited by editor
% \dates{Compiled \today}

%\ociscodes{(140.3490) Lasers, distributed feedback; (060.2420) Fibers, polarization-maintaining; (060.3735) Fiber Bragg gratings.}

% To be edited by editor
% \doi{\url{http://dx.doi.org/10.1364/optica.XX.XXXXXX}}

\begin{abstract}
We describe theoretically and verify experimentally a novel class of diffraction-free pulsed optical beams that are `omni-resonant': they have the remarkable property of transmission through planar Fabry-P{\'e}rot resonators without spectral filtering even if their bandwidth far exceeds the cavity resonant linewidth. Ultrashort wave packets endowed with a specific spatio-temporal structure couple to a \textit{single} resonant mode independently of its linewidth. We confirm that such `space-time' omni-resonant wave packets retain their bandwidth (1.6~nm), spatio-temporal profile (1.3-ps pulse width, 4-$\mu$m beam width), and diffraction-free behavior upon transmission through cavities with resonant linewidths of 0.3-nm and 0.15-nm.
\end{abstract}

\maketitle

Space-time (ST) wave packets constitute a broad family of pulsed optical beams endowed with a distinct feature: each spatial frequency is tightly associated with a particular wavelength \cite{Donnelly93PRSLA,Saari04PRE,Longhi04OE,Kondakci16OE,Parker16OE}. In contrast, such an association is absent from traditional wave packets where the spatial and temporal spectra are typically independent of each other. By introducing precise spatio-temporal spectral structure into ST wave packets, unique and surprising characteristics emerge, such as propagation-invariant propagation \cite{Brittingham83JAP,Lu92IEEEa}, arbitrary group velocities in free space \cite{Salo01JOA,Saari04PRE,Longhi04OE}, dispersion cancellation in optical materials \cite{Sonajalg97OL,Porras01OL}, among other possibilities \cite{Faccio06PRL,Porras17OL,Efremidis17OL,Wong17ACSP2}. Such wave packets have been studied theoretically for more than three decades \cite{Reivelt03arxiv,Kiselev07OS,Turunen10PO,FigueroaBook14}, and particular examples have been realized experimentally \cite{Saari97PRL,Reivelt00JOSAA,Reivelt02PRE}. We have recently introduced a novel phase-only approach to the synthesis of ST wave packets \cite{Kondakci17NP,Yessenov19OPN} that enables precise control over the properties of ST wave packets, leading to the observation of arbitrary group velocities in free space (from $30c$ to $-4c$) \cite{Kondakci19NC} and in optical materials \cite{Bhaduri19Optica}, self-healing \cite{Kondakci18OL}, non-accelerating Airy wave packets \cite{Kondakci18PRL}, and corresponding effects realized using incoherent light \cite{Yessenov19Optica,Yessenov19OL}. It is an open question how the spatio-temporal structure of such wave packets affects their interaction with photonic devices.

\begin{figure}[b!]
  \begin{center}
  \includegraphics[width=8.6cm]{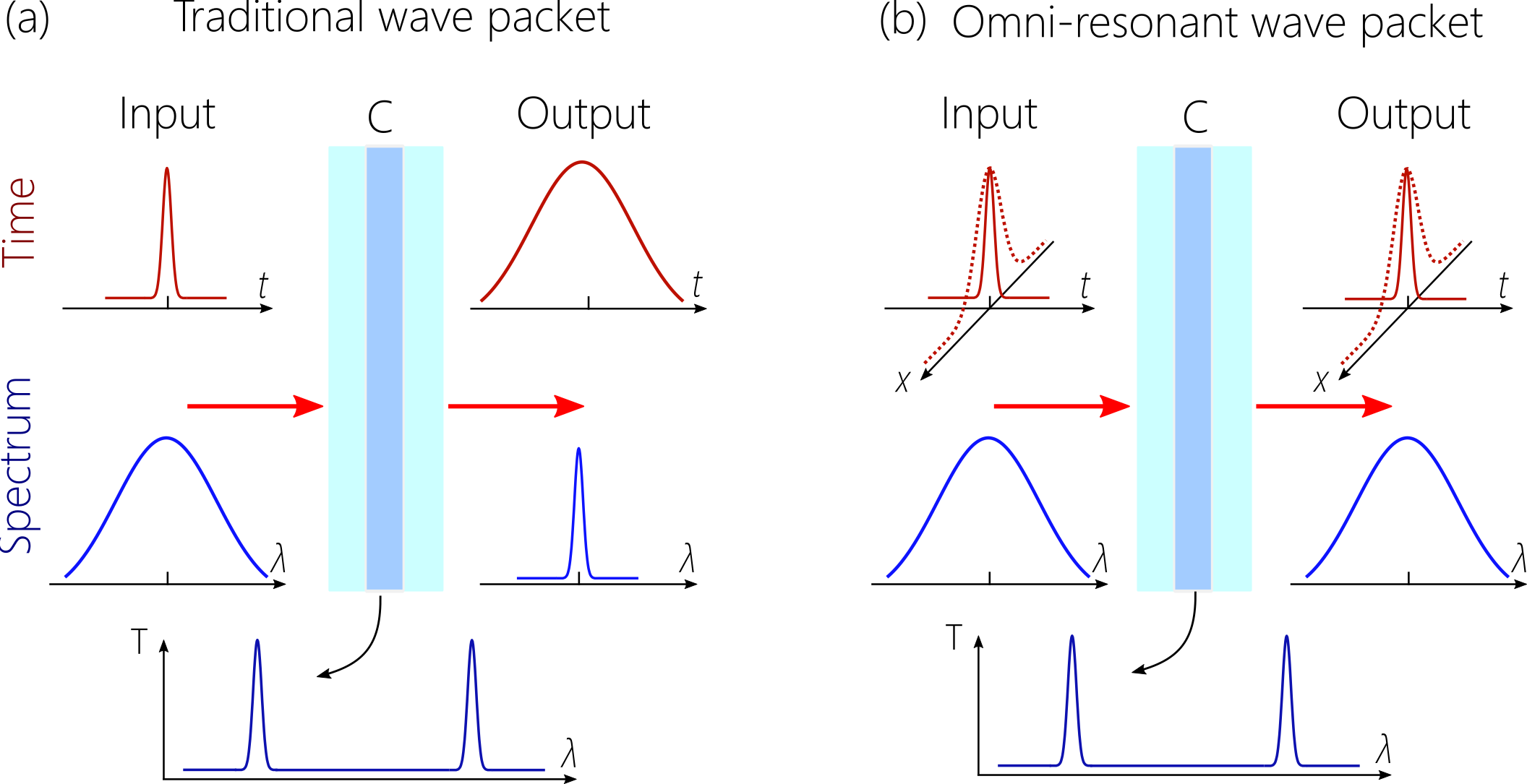}
  \end{center}
  \caption{Concept of omni-resonant wave packets. (a) A traditional wave packet is spectrally filtered and temporally broadened when traversing a FP cavity whose linewidth is narrower than its bandwidth. C: FP cavity; T: spectral transmission. (b) An omni-resonant wave packet is transmitted through a FP cavity unchanged regardless of its bandwidth.}
  \label{Fig:Concept}
\end{figure}

Here we show that tailoring the spatio-temporal spectral structure of a ST wave packet can render it `omni-resonant': it can traverse a planar Fabry-P{\'e}rot (FP) resonator \textit{without} spectral filtering -- even if the wave-packet bandwidth is larger than the cavity resonant linewidth. Critically, the entire wave-packet spectrum couples to the \textit{same} resonant mode. We demonstrate this surprising behavior using wave packets of bandwidth 1.6~nm interacting with FP cavities having resonant linewidths of 0.3~nm and 0.15~nm at a wavelength of $\sim\!800$~nm. The spatio-temporal profiles and the diffraction-free behavior of omni-resonant ST wave packets remain unchanged after traversing the cavities. These results may lead to novel \textit{resonantly} enhanced -- and yet spectrally \textit{broadband} -- nonlinear optical processes, and may potentially enhance solar-energy harvesting \cite{Villinger19unpubl}.

\begin{figure*}[t!]
  \begin{center}
  \includegraphics[width=17.2cm]{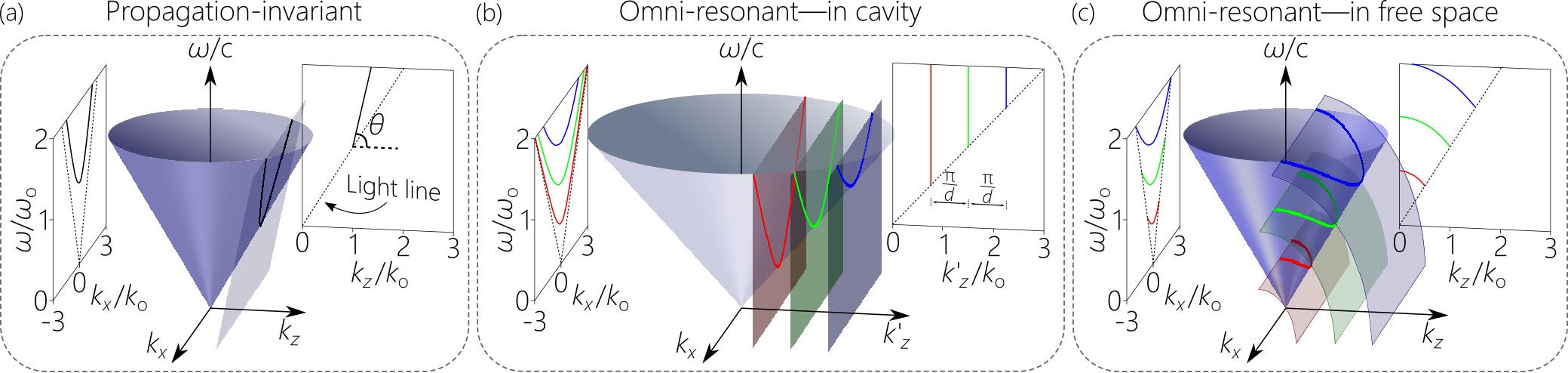}
  \end{center}
  \caption{Spatio-temporal spectral representations of ST wave packets. (a) Propagation-invariant ST wave packets lie along the conic section at the intersection of the free-space light-cone $k_{x}^{2}+k_{z}^{2}\!=\!(\tfrac{\omega}{c})^{2}$ with a tilted spectral plane making an angle $\theta$ with respect to the $k_{z}$-axis. (b) Resonant modes of a FP cavity lie along the hyperbola at the intersection of the cavity-material light-cone $k_{x}^{2}+k_{z}'^{2}\!=\!(n\tfrac{\omega}{c})^{2}$ with vertical iso-$k_{z}'$ planes, $k_{z}'\!=\!k_{m}\!=\!m\tfrac{\pi}{d}$. (c) Transforming the spectral representation in (b) onto the free-space light-cone while maintaining the invariance of $k_{x}$ and $\omega$ to find the free-space spectral representation of omni-resonant ST wave packets.}
  \label{Fig:Representations}
\end{figure*}

A planar FP cavity transmits light within a narrow spectral linewidth (inversely proportional to the cavity-photon lifetime) centered at the resonances. A narrow pulse traversing the cavity broadens in time after spectral filtering when its bandwidth exceeds the cavity linewidth [Fig.~\ref{Fig:Concept}(a)]. An omni-resonant wave packet, on the other hand, is transmitted through the same cavity without spectral filtering or temporal broadening \textit{even if} its bandwidth exceeds the cavity linewidth by virtue of the wave packet structure [Fig.~\ref{Fig:Concept}(b)]. Our work is therefore distinct from so-called `white-light cavities' \cite{Wicht97OC} where a resonance is spectrally broadened by incorporating into the cavity a medium that provides anomalous dispersion (e.g., via Raman gain in an atomic vapor \cite{Pati07PRL}, electromagnetically induced transparency \cite{Wu08PRA}, or nonlinear Brillouin scattering \cite{Yum13JLT}). In fact, previous research has established the \textit{impossibility} of producing this effect via linear intra-cavity components \cite{Wise05PRL,Yum13OC}. In contrast, we do \textit{not} modify the cavity, and omni-resonance instead results from a specific spatio-temporal structure introduced into the optical wave packet -- whereas previous investigations of white-light cavities have ignored the transverse spatial structure of the field. We previously demonstrated omni-resonance using broadband incoherent light by introducing angular dispersion into the wave front \cite{Shabahang17SR,Shabahang19OL}, and have exploited this in boosting the photocurrent harvested by a cavity-embedded solar cell in the near-infrared \cite{Villinger19unpubl}. In contrast, we show here that \textit{diffraction-free} ST wave packets can resonate in a cavity and yet retain their diffraction-free behavior, transverse spatial size, temporal profile, and bandwidth independently of the resonance linewidth.

We first examine propagation-invariant ST wave packets in free space \cite{Kondakci17NP,Yessenov19OPN} whose spatio-temporal spectral representations lie along the intersection of the light-cone $k_{x}^{2}+k_{z}^{2}\!=\!(\tfrac{\omega}{c})^{2}$ with tilted spectral planes described by the equation $\tfrac{\omega}{c}\!=\!k_{\mathrm{o}}+(k_{z}-k_{\mathrm{o}})\tan{\theta}$; here $x$ and $k_{x}$ are the transverse coordinate and component of the wave vector, respectively, $z$ and $k_{z}$ are their axial counterparts, $\omega$ is the (temporal) frequency, $c$ is the speed of light in vacuum, $k_{\mathrm{o}}$ is a fixed wave number ($k_{z}\!=\!k_{\mathrm{o}}\!=\!\tfrac{\omega_{\mathrm{o}}}{c}$ when $k_{x}\!=\!0$), $\theta$ is the spectral tilt angle with respect to the $k_{z}$-axis, and we hold the field uniform along $y$ ($k_{y}\!=\!0$); see Fig.~\ref{Fig:Representations}(a). The critical feature to note is that each spatial frequency $k_{x}$ is associated with a single temporal frequency $\omega$. Here the spectral projection onto the $(k_{z},\tfrac{\omega}{c})$-plane is a straight line, indicating that the wave packet travels at an axial group velocity $v_{\mathrm{g}}\!=\!\tfrac{\partial\omega}{\partial k_{z}}\!=\!c\tan{\theta}$ and the wave packet envelope is transported rigidly in free space, $\psi(x,z,t)\!=\!\psi(x,0,t-z/v_{\mathrm{g}})$.

Now consider a FP cavity consisting of a layer of index $n$ and thickness $d$ placed between two symmetric mirrors. Only plane waves whose axial wave number in the cavity $k'_{z}$ belongs to a discrete set $k_{m}\!=\!m\pi/d$ resonate, where $m$ is the modal index (we place a prime on only $k_{z}$ because $\omega$ and $k_{x}$ are invariant). All such plane waves associated with the $m^{\mathrm{th}}$-order resonance lie along the intersection of the cavity-material light-cone $k_{x}^{2}+k_{z}'^{2}\!=\!(n\tfrac{\omega}{c})^{2}$ with vertical iso-$k_{z}'$-plane ($k_{z}'\!=\!k_{m}$; i.e., $\theta\!=\!90^{\circ}$), which takes the form of a hyperbola [Fig.~\ref{Fig:Representations}(b)],
\begin{equation}\label{Eq:kxOmegaHyperbola}
%n^{2}(\omega/c)^{2}-k_{x}^{2}=m^{2}(\pi/d)^{2}.
n^{2}\left(\frac{\omega}{c}\right)^{2}-k_{x}^{2}=m^{2}\left(\frac{\pi}{d}\right)^{2}.
\end{equation}
Critically, each $k_{x}$ is associated with a single $\omega$. Therefore, a broadband pulse can resonate in its entirety with a cavity resonance even if its bandwidth exceeds the cavity linewidth as long as its \textit{spatio-temporal spectrum} conforms to the hyperbola in Eq.~\ref{Eq:kxOmegaHyperbola}.

The spectral representation in Fig.~\ref{Fig:Representations}(b) is for the field \textit{inside} the cavity. To couple a field from free space into the cavity, we transform this spectrum onto the free-space light-cone constrained by the invariance of $\omega$ and $k_{x}$; that is, while retaining the spectral projection onto the $(k_{x},\tfrac{\omega}{c})$-plane to be that in Eq.~\ref{Eq:kxOmegaHyperbola} \cite{Bhaduri19Optica}. Such a transformation, however, changes the projection onto the $(k_{z},\tfrac{\omega}{c})$-plane from a straight line in the cavity [Fig.~\ref{Fig:Representations}(b)] into a section of an ellipse in free space [Fig.~\ref{Fig:Representations}(c)], 
\begin{equation}\label{Eq:kzOmegaProjection}
%(n^{2}-1)(\omega/c)^{2}+k_{z}^{2}=m^{2}(\pi/d)^{2}.
(n^{2}-1)\left(\frac{\omega}{c}\right)^{2}+k_{z}^{2}=m^{2}\left(\frac{\pi}{d}\right)^{2}.
\end{equation}
In other words, the spatio-temporal spectrum of the wave packet in free space lies along the intersection of the light-cone with the curved surface described by Eq.~\ref{Eq:kzOmegaProjection}. Because the projection onto the $(k_{z},\tfrac{\omega}{c})$-plane is no longer a straight line, the wave packet undergoes group velocity dispersion in free space (but \textit{not} in the cavity), with group velocity $v_{\mathrm{g}}\!=\!-\tfrac{1}{n^{2}-1}c$ and dispersion coefficient $k_{2}\!=\!\tfrac{\partial^{2}k_{z}}{\partial\omega^{2}}\!=\!-\tfrac{n^{2}}{n^{2}-1}c^{2}k_{\mathrm{o}}$, so that $v_{\mathrm{g}}$ in free space is negative. Negative-$v_{\mathrm{g}}$ ST wave packets have been studied theoretically \cite{Zapata06OL} and realized experimentally \cite{Kondakci19NC}, and do not violate relativistic causality \cite{Shaarawi00JPA,SaariPRA18,Yessenov19OE}

\begin{figure}[t!]
  \begin{center}
  \includegraphics[width=8.6cm]{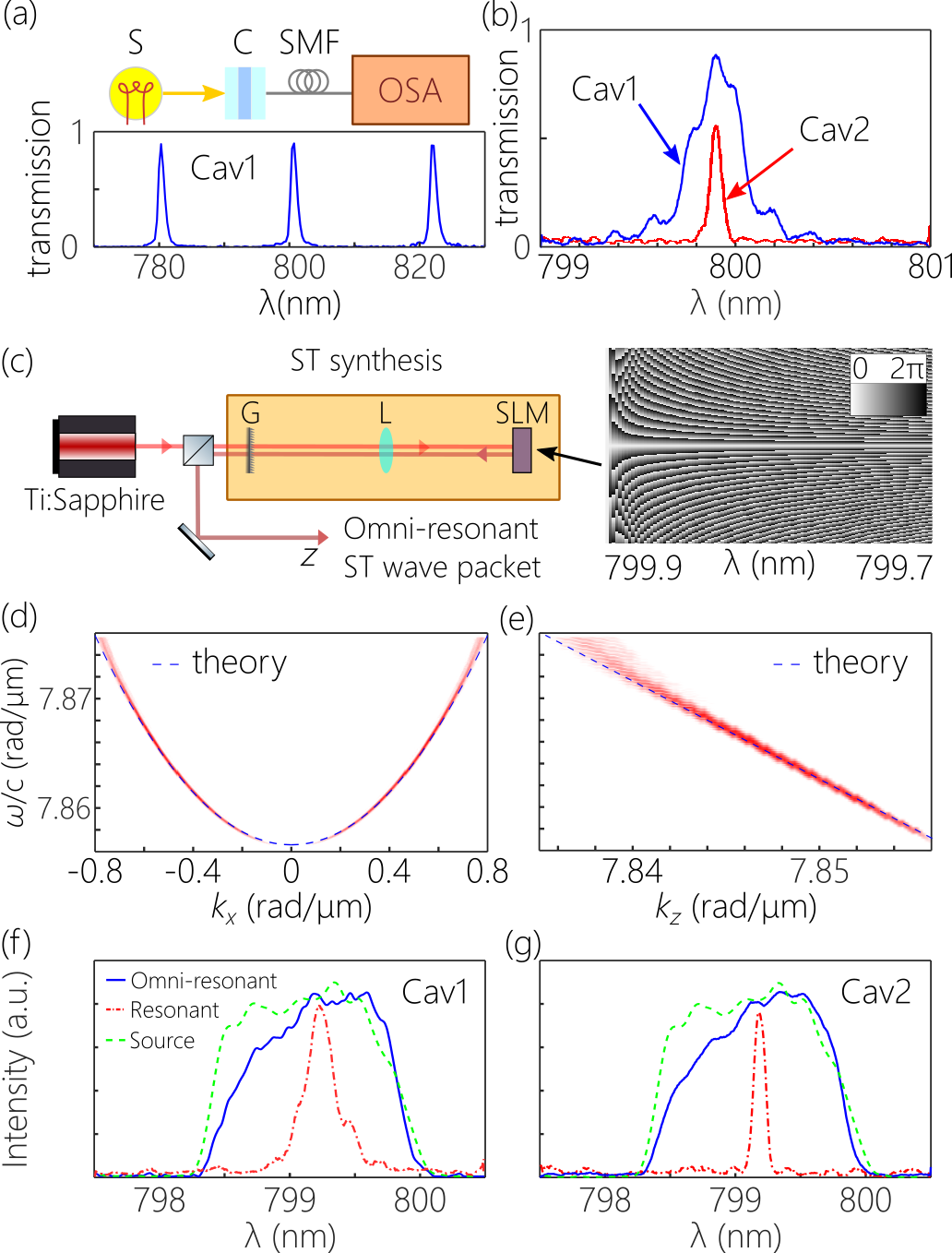}
  \end{center}
  \caption{{(a) Transmitted spectrum through Cav1. The experimental setup is sketched on top. S: Optical source (halogen lamp); SMF: single-mode fiber; OSA: optical spectrum analyzer. (b) Comparison of the resonant linewidth of Cav1 and Cav2. (c) Schematic of the optical setup for synthesizing omni-resonant ST wave packets. G: Diffraction grating; L: cylindrical lens; SLM: spatial light modulator. Diagram to the right depicts the phase pattern imparted by the SLM to the incident wave front to render the wave packet omni-resonant. (d) Measured spatio-temporal spectrum in the $(k_{x},\tfrac{\omega}{c})$ and (e) $(k_{z},\tfrac{\omega}{c})$ domains. Dashed lines are theoretical predictions based on Eqs.~\ref{Eq:kxOmegaHyperbola},\ref{Eq:kzOmegaProjection}. (f) Transmitted spectrum of the omni-resonant ST wave packet for Cav1 and (g) for Cav2, which are substantially broader than the resonant linewidth of each cavity and approach that of the as-synthesized wave packet.}}
  \label{Fig:Setup}
\end{figure}

We carried out our experiments using two FP cavities, each formed of symmetric Bragg mirrors sandwiching a 10-$\mu$m-thick layer of SiO$_2$ (index of $n\!=\!1.45$ at a wavelength of $\lambda\!=\!800$~nm, leading to a free spectral range of $\sim\!22$~nm) deposited via e-beam evaporation. The Bragg mirrors are formed of bilayers of SiO$_2$ and TiO$_2$ (thicknesses of 138~nm and 88~nm, and indices at $\lambda\!=\!800$~nm of 1.45 and 2.28, respectively). One cavity (referred to hereon as Cav1) comprises mirrors containing 6 bilayers and the other (Cav2) 8 bilayers, leading to linewidths of 0.3~nm (Cav1) and 0.15~nm (Cav2), as determined by a collimated beam from a halogen lamp (Thorlabs QTH10) [Fig.~\ref{Fig:Setup}(a,b)].

We synthesized omni-resonant ST wave packets utilizing the setup in Fig.~\ref{Fig:Setup}(c) \cite{Kondakci17NP,Yessenov19OPN}. Pulses from a Ti:Sapphire laser centered at $\sim\!800$~nm are directed to a diffraction grating (Newport 10HG1200-800-1, 1200 lines/mm, area $25\!\times\!25$~mm$^{2}$) that spreads the spectrum in space, and a cylindrical lens (focal length $f\!=\!500$~mm) collimates the spectrum and directs it to a spatial light modulator (SLM; Hamamatsu X10468-02). The SLM assigns to each wavelength a linear phase distribution corresponding to a prescribed spatial frequency $k_{x}$ according to the hyperbolic spatio-temporal relationship in Eq.~\ref{Eq:kxOmegaHyperbola} via the phase distribution shown in Fig.~\ref{Fig:Setup}(c). The phase-modulated wave front is retro-reflected back to the grating that reconstitutes the pulse and produces the ST wave packet, and two lenses then introduce a demagnification factor of $10\times$ to compensate for spatial scaling at the SLM. Such a wave packet will be diffraction-free for a distance dictated by the `spectral uncertainty' \cite{Yessenov19OE}, which is determined by the spectral resolution of the grating. To confirm that the desired spatio-temporal spectral structure has indeed been introduced into the wave packet, we measure simultaneously the spectral intensity after resolving the wavelengths via a grating and the spatial spectrum via a Fourier transforming lens. The measurements in the $(k_{x},\tfrac{\omega}{c})$-plane are plotted in Fig.~\ref{Fig:Setup}(d), corresponding to Eq.~\ref{Eq:kxOmegaHyperbola}, and in the $(k_{z},\tfrac{\omega}{c})$-plane in Fig.~\ref{Fig:Setup}(e), corresponding to Eq.~\ref{Eq:kzOmegaProjection}, with excellent agreement between the measurements and the theoretical expectations.

\begin{figure}[t!]
  \begin{center}
  \includegraphics[width=8.6cm]{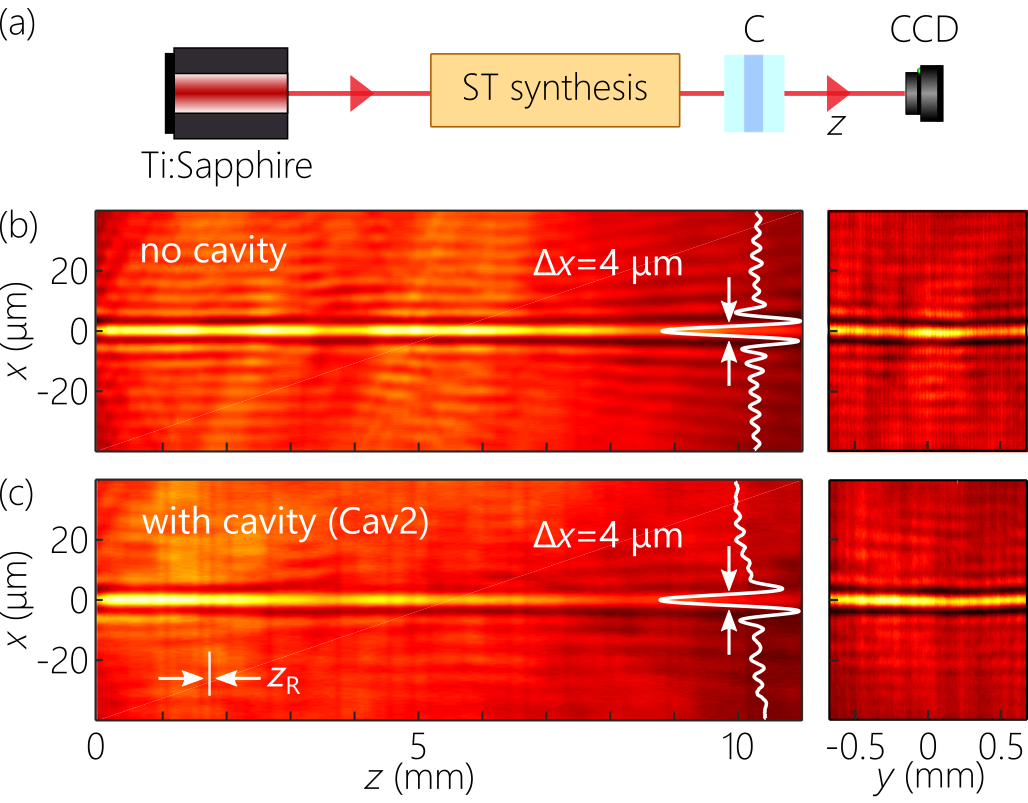}
  \end{center}
  \caption{(a) Setup for measuring the axial evolution of the omni-resonant ST wave packet. C: Cavity. (b) Diffraction-free propagation of the omni-resonant ST wave packet in absence and (c) in presence of the FP cavity. The thin white line represents the Rayleigh range of a Gaussian beam having the same transverse width of the omni-resonant ST wave packet. On the right we plot the intensity in the transverse plane at $z\!=\!0$.}
  \label{Fig:DiffractionFree}
\end{figure}

The omni-resonance of this ST wave packet is confirmed by measuring the wave packet spectra after transmission of the same ST wave packet through Cav1 [Fig.~\ref{Fig:Setup}(f)] and Cav2 [Fig.~\ref{Fig:Setup}(g)], and comparing them to the spectrum of the as-synthesized wave packet. We further compare these spectra to the resonance linewidth of the bare cavity and verify that for both cavities the transmitted spectra of the omni-resonant ST wave packets match that of the synthesized wave packet despite being larger than the resonant linewidths. Next, we confirm that the ST wave packet retains its diffraction-free characteristics after transmission through the cavity by measuring the time-averaged intensity along the propagation axis $I(x,z)$ with a CCD camera (ImagingSource DMK 27BUP031) that is scanned along $z$ [Fig.~\ref{Fig:DiffractionFree}(a)]. The measurements confirm that the axial evolution and the size and shape of the spatial profile of the omni-resonant ST wave packet in \textit{absence} of a FP cavity [Fig.~\ref{Fig:DiffractionFree}(b)] are maintained in \textit{presence} of a FP cavity (Cav2) [Fig.~\ref{Fig:DiffractionFree}(c)].

\begin{figure}[t!]
  \begin{center}
  \includegraphics[width=8.6cm]{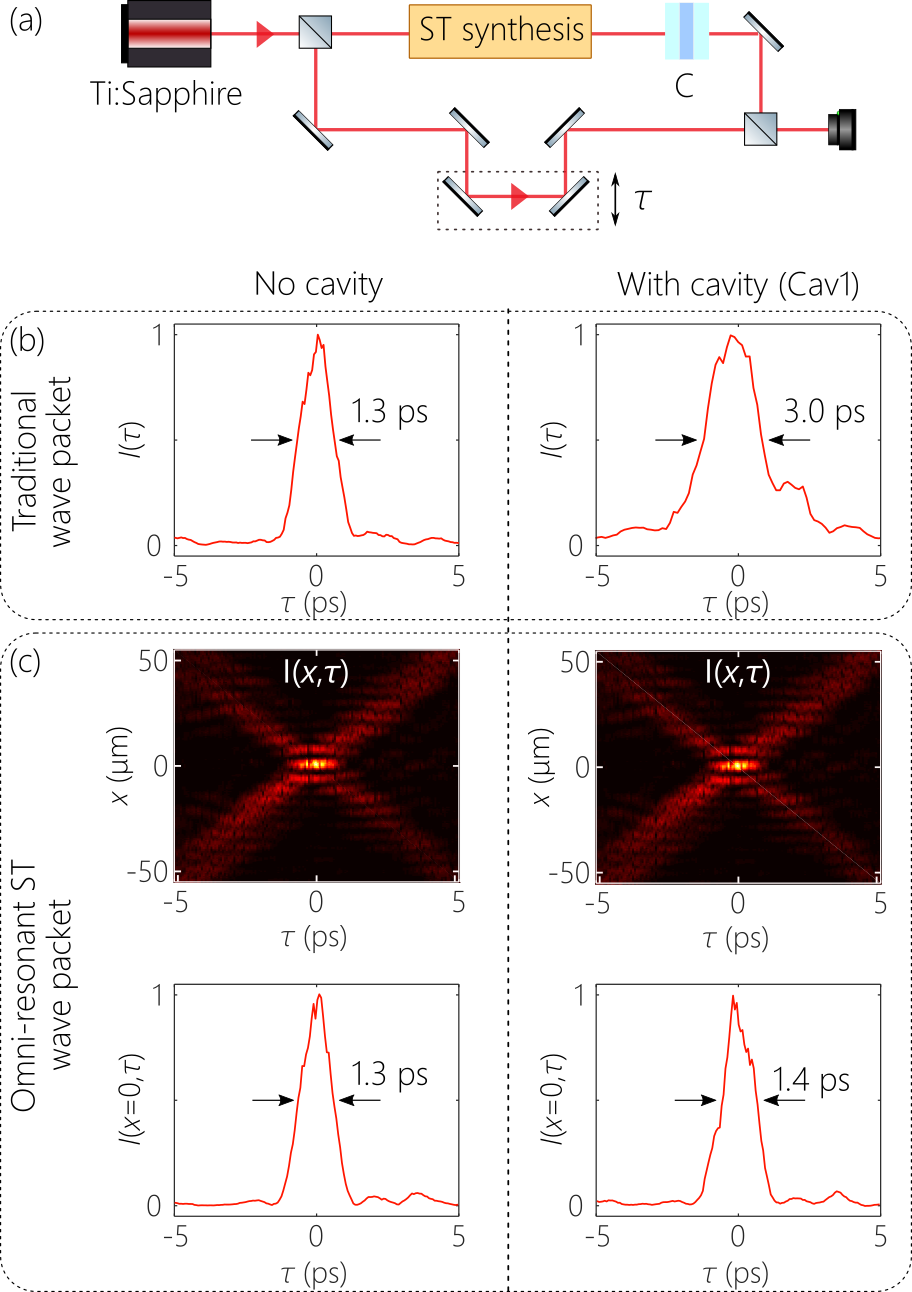}
  \end{center}
  \caption{{(a) Setup for tracing the spatio-temporal intensity profile of the omni-resonant ST wave packets. (b) Temporal profiles before and after a FP cavity (Cav1) for a traditional wave packet. (c) Measured spatio-temporal intensity profiles before and after the FP cavity (Cav1) demonstrating omni-resonance.}}
  \label{Fig:TimeResolved}
\end{figure}

Finally, we examine the pulse width of omni-resonance ST wave packets upon traversing a cavity. To trace the spatio-temporal intensity profile of the wave packet, we embed the synthesis arrangement [Fig.~\ref{Fig:Setup}(c)] into a Mach-Zehnder interferometer [Fig.~\ref{Fig:TimeResolved}(a)]. The initial femtosecond pulse is directed to a reference arm where it undergoes an optical delay $\tau$, and is then brought together with the synthesized ST wave packet. When the two wave packets overlap in space and time, spatially resolved fringes are observed whose visibility allows the extraction of the spatio-temporal profile of the ST wave packet envelope as the delay is swept \cite{Kondakci19NC,Bhaduri19Optica}.

We first record the pulse envelopes for a traditional wave packet before and after transmission through the cavity [Fig.~\ref{Fig:TimeResolved}(b)]. This measurement is carried out by simply idling the SLM in the synthesis arrangement. The pulse width increases from 1.3~ps to 3~ps after traversing Cav1 as the 1.6-nm-bandwidth pulse spectrum is reduced to 0.3~nm. When repeating this measurement after activating the SLM to synthesize the omni-resonant ST wave packet, an altogether different behavior emerges. The spatio-temporal profile $I(x,\tau)$ at fixed $z$ of the ST wave packet is shown in Fig.~\ref{Fig:TimeResolved}(c) displaying the X-shape characteristic of such pulsed beams where tails extend away from the wave packet center. We find that this profile is unchanged after traversing the cavity (just as its spectrum is invariant [Fig.~\ref{Fig:Setup}(f)]. Sections along $x\!=\!0$ of the wave-packet central feature, $I(0,\tau)$ confirm that the temporal profile of the omni-resonant ST wave packet is unaffected by transmission through the cavity.

In conclusion, we have demonstrated a general methodology for synthesizing \textit{omni-resonant wave packets} through the introduction of specific spatio-temporal structure into the optical field. This approach allows the advantages accrued \textit{on} resonance in a planar cavity (such as field build up) to be harnessed \textit{over a large bandwidth} rather than only within the resonant linewidth \cite{Villinger19unpubl}. This unique feature can have applications in broadband resonantly enhanced nonlinear effects. Note that so-called `baseband' ST wave packets whose spatial spectrum can extend down to $k_{x}\!=\!0$ \cite{Yessenov19PRA}, such as those studied here, are necessary for observing omni-resonant behaviour, whereas X-waves \cite{Lu92IEEEa,Saari97PRL} and focus-wave modes \cite{Brittingham83JAP,Reivelt00JOSAA,Reivelt02PRE} can\textit{not} display omni-resonance. Finally, this work points to a broader utility of ST wave packets in tailoring interactions with photonic devices through arbitrary control exercised over the spatio-temporal structure of light. 

\section*{Funding}
U.S. Office of Naval Research (ONR) contract N00014-17-1-2458.

%\vspace{2mm}
%\noindent
%\textbf{Disclosures.} The authors declare no conflicts of interest.

\bibliography{diffraction}

\begin{thebibliography}{43}%
\makeatletter
\providecommand \@ifxundefined [1]{%
 \@ifx{#1\undefined}
}%
\providecommand \@ifnum [1]{%
 \ifnum #1\expandafter \@firstoftwo
 \else \expandafter \@secondoftwo
 \fi
}%
\providecommand \@ifx [1]{%
 \ifx #1\expandafter \@firstoftwo
 \else \expandafter \@secondoftwo
 \fi
}%
\providecommand \natexlab [1]{#1}%
\providecommand \enquote  [1]{``#1''}%
\providecommand \bibnamefont  [1]{#1}%
\providecommand \bibfnamefont [1]{#1}%
\providecommand \citenamefont [1]{#1}%
\providecommand \href@noop [0]{\@secondoftwo}%
\providecommand \href [0]{\begingroup \@sanitize@url \@href}%
\providecommand \@href[1]{\@@startlink{#1}\@@href}%
\providecommand \@@href[1]{\endgroup#1\@@endlink}%
\providecommand \@sanitize@url [0]{\catcode `\\12\catcode `\$12\catcode
  `\&12\catcode `\#12\catcode `\^12\catcode `\_12\catcode `\%12\relax}%
\providecommand \@@startlink[1]{}%
\providecommand \@@endlink[0]{}%
\providecommand \url  [0]{\begingroup\@sanitize@url \@url }%
\providecommand \@url [1]{\endgroup\@href {#1}{\urlprefix }}%
\providecommand \urlprefix  [0]{URL }%
\providecommand \Eprint [0]{\href }%
\providecommand \doibase [0]{http://dx.doi.org/}%
\providecommand \selectlanguage [0]{\@gobble}%
\providecommand \bibinfo  [0]{\@secondoftwo}%
\providecommand \bibfield  [0]{\@secondoftwo}%
\providecommand \translation [1]{[#1]}%
\providecommand \BibitemOpen [0]{}%
\providecommand \bibitemStop [0]{}%
\providecommand \bibitemNoStop [0]{.\EOS\space}%
\providecommand \EOS [0]{\spacefactor3000\relax}%
\providecommand \BibitemShut  [1]{\csname bibitem#1\endcsname}%
\let\auto@bib@innerbib\@empty
%</preamble>
\bibitem [{\citenamefont {Donnelly}\ and\ \citenamefont
  {Ziolkowski}(1993)}]{Donnelly93PRSLA}%
  \BibitemOpen
  \bibfield  {author} {\bibinfo {author} {\bibfnamefont {R.}~\bibnamefont
  {Donnelly}}\ and\ \bibinfo {author} {\bibfnamefont {R.}~\bibnamefont
  {Ziolkowski}},\ }\href@noop {} {\bibfield  {journal} {\bibinfo  {journal}
  {Proc. R. Soc. Lond. A}\ }\textbf {\bibinfo {volume} {440}},\ \bibinfo
  {pages} {541} (\bibinfo {year} {1993})}\BibitemShut {NoStop}%
\bibitem [{\citenamefont {Saari}\ and\ \citenamefont
  {Reivelt}(2004)}]{Saari04PRE}%
  \BibitemOpen
  \bibfield  {author} {\bibinfo {author} {\bibfnamefont {P.}~\bibnamefont
  {Saari}}\ and\ \bibinfo {author} {\bibfnamefont {K.}~\bibnamefont
  {Reivelt}},\ }\href@noop {} {\bibfield  {journal} {\bibinfo  {journal} {Phys.
  Rev. E}\ }\textbf {\bibinfo {volume} {69}},\ \bibinfo {pages} {036612}
  (\bibinfo {year} {2004})}\BibitemShut {NoStop}%
\bibitem [{\citenamefont {Longhi}(2004)}]{Longhi04OE}%
  \BibitemOpen
  \bibfield  {author} {\bibinfo {author} {\bibfnamefont {S.}~\bibnamefont
  {Longhi}},\ }\href@noop {} {\bibfield  {journal} {\bibinfo  {journal} {Opt.
  Express}\ }\textbf {\bibinfo {volume} {12}},\ \bibinfo {pages} {935}
  (\bibinfo {year} {2004})}\BibitemShut {NoStop}%
\bibitem [{\citenamefont {Kondakci}\ and\ \citenamefont
  {Abouraddy}(2016)}]{Kondakci16OE}%
  \BibitemOpen
  \bibfield  {author} {\bibinfo {author} {\bibfnamefont {H.~E.}\ \bibnamefont
  {Kondakci}}\ and\ \bibinfo {author} {\bibfnamefont {A.~F.}\ \bibnamefont
  {Abouraddy}},\ }\href@noop {} {\bibfield  {journal} {\bibinfo  {journal}
  {Opt. Express}\ }\textbf {\bibinfo {volume} {24}},\ \bibinfo {pages} {28659}
  (\bibinfo {year} {2016})}\BibitemShut {NoStop}%
\bibitem [{\citenamefont {Parker}\ and\ \citenamefont
  {Alonso}(2016)}]{Parker16OE}%
  \BibitemOpen
  \bibfield  {author} {\bibinfo {author} {\bibfnamefont {K.~J.}\ \bibnamefont
  {Parker}}\ and\ \bibinfo {author} {\bibfnamefont {M.~A.}\ \bibnamefont
  {Alonso}},\ }\href@noop {} {\bibfield  {journal} {\bibinfo  {journal} {Opt.
  Express}\ }\textbf {\bibinfo {volume} {24}},\ \bibinfo {pages} {28669}
  (\bibinfo {year} {2016})}\BibitemShut {NoStop}%
\bibitem [{\citenamefont {Brittingham}(1983)}]{Brittingham83JAP}%
  \BibitemOpen
  \bibfield  {author} {\bibinfo {author} {\bibfnamefont {J.~N.}\ \bibnamefont
  {Brittingham}},\ }\href@noop {} {\bibfield  {journal} {\bibinfo  {journal}
  {J. Appl. Phys.}\ }\textbf {\bibinfo {volume} {54}},\ \bibinfo {pages} {1179}
  (\bibinfo {year} {1983})}\BibitemShut {NoStop}%
\bibitem [{\citenamefont {Lu}\ and\ \citenamefont
  {Greenleaf}(1992)}]{Lu92IEEEa}%
  \BibitemOpen
  \bibfield  {author} {\bibinfo {author} {\bibfnamefont {J.-Y.}\ \bibnamefont
  {Lu}}\ and\ \bibinfo {author} {\bibfnamefont {J.~F.}\ \bibnamefont
  {Greenleaf}},\ }\href@noop {} {\bibfield  {journal} {\bibinfo  {journal}
  {IEEE Trans. Ultrason. Ferroelec. Freq. Control}\ }\textbf {\bibinfo {volume}
  {39}},\ \bibinfo {pages} {19} (\bibinfo {year} {1992})}\BibitemShut {NoStop}%
\bibitem [{\citenamefont {Salo}\ and\ \citenamefont
  {Salomaa}(2001)}]{Salo01JOA}%
  \BibitemOpen
  \bibfield  {author} {\bibinfo {author} {\bibfnamefont {J.}~\bibnamefont
  {Salo}}\ and\ \bibinfo {author} {\bibfnamefont {M.~M.}\ \bibnamefont
  {Salomaa}},\ }\href@noop {} {\bibfield  {journal} {\bibinfo  {journal} {J.
  Opt. A}\ }\textbf {\bibinfo {volume} {3}},\ \bibinfo {pages} {366} (\bibinfo
  {year} {2001})}\BibitemShut {NoStop}%
\bibitem [{\citenamefont {S{\~o}najalg}\ \emph {et~al.}(1997)\citenamefont
  {S{\~o}najalg}, \citenamefont {R{\"a}tsep},\ and\ \citenamefont
  {Saari}}]{Sonajalg97OL}%
  \BibitemOpen
  \bibfield  {author} {\bibinfo {author} {\bibfnamefont {H.}~\bibnamefont
  {S{\~o}najalg}}, \bibinfo {author} {\bibfnamefont {M.}~\bibnamefont
  {R{\"a}tsep}}, \ and\ \bibinfo {author} {\bibfnamefont {P.}~\bibnamefont
  {Saari}},\ }\href@noop {} {\bibfield  {journal} {\bibinfo  {journal} {Opt.
  Lett.}\ }\textbf {\bibinfo {volume} {22}},\ \bibinfo {pages} {310} (\bibinfo
  {year} {1997})}\BibitemShut {NoStop}%
\bibitem [{\citenamefont {Porras}(2001)}]{Porras01OL}%
  \BibitemOpen
  \bibfield  {author} {\bibinfo {author} {\bibfnamefont {M.~A.}\ \bibnamefont
  {Porras}},\ }\href@noop {} {\bibfield  {journal} {\bibinfo  {journal} {Opt.
  Lett.}\ }\textbf {\bibinfo {volume} {26}},\ \bibinfo {pages} {1364} (\bibinfo
  {year} {2001})}\BibitemShut {NoStop}%
\bibitem [{\citenamefont {Faccio}\ \emph {et~al.}(2006)\citenamefont {Faccio},
  \citenamefont {Porras}, \citenamefont {Dubietis}, \citenamefont {Bragheri},
  \citenamefont {Couairon},\ and\ \citenamefont {{Di T}rapani}}]{Faccio06PRL}%
  \BibitemOpen
  \bibfield  {author} {\bibinfo {author} {\bibfnamefont {D.}~\bibnamefont
  {Faccio}}, \bibinfo {author} {\bibfnamefont {M.~A.}\ \bibnamefont {Porras}},
  \bibinfo {author} {\bibfnamefont {A.}~\bibnamefont {Dubietis}}, \bibinfo
  {author} {\bibfnamefont {F.}~\bibnamefont {Bragheri}}, \bibinfo {author}
  {\bibfnamefont {A.}~\bibnamefont {Couairon}}, \ and\ \bibinfo {author}
  {\bibfnamefont {P.}~\bibnamefont {{Di T}rapani}},\ }\href@noop {} {\bibfield
  {journal} {\bibinfo  {journal} {Phys. Rev. Lett.}\ }\textbf {\bibinfo
  {volume} {96}},\ \bibinfo {pages} {193901} (\bibinfo {year}
  {2006})}\BibitemShut {NoStop}%
\bibitem [{\citenamefont {Porras}(2017)}]{Porras17OL}%
  \BibitemOpen
  \bibfield  {author} {\bibinfo {author} {\bibfnamefont {M.~A.}\ \bibnamefont
  {Porras}},\ }\href@noop {} {\bibfield  {journal} {\bibinfo  {journal} {Opt.
  Lett.}\ }\textbf {\bibinfo {volume} {42}},\ \bibinfo {pages} {4679} (\bibinfo
  {year} {2017})}\BibitemShut {NoStop}%
\bibitem [{\citenamefont {Efremidis}(2017)}]{Efremidis17OL}%
  \BibitemOpen
  \bibfield  {author} {\bibinfo {author} {\bibfnamefont {N.~K.}\ \bibnamefont
  {Efremidis}},\ }\href@noop {} {\bibfield  {journal} {\bibinfo  {journal}
  {Opt. Lett.}\ }\textbf {\bibinfo {volume} {42}},\ \bibinfo {pages} {5038}
  (\bibinfo {year} {2017})}\BibitemShut {NoStop}%
\bibitem [{\citenamefont {Wong}\ and\ \citenamefont
  {Kaminer}(2017)}]{Wong17ACSP2}%
  \BibitemOpen
  \bibfield  {author} {\bibinfo {author} {\bibfnamefont {L.~J.}\ \bibnamefont
  {Wong}}\ and\ \bibinfo {author} {\bibfnamefont {I.}~\bibnamefont {Kaminer}},\
  }\href@noop {} {\bibfield  {journal} {\bibinfo  {journal} {ACS Photon.}\
  }\textbf {\bibinfo {volume} {4}},\ \bibinfo {pages} {2257} (\bibinfo {year}
  {2017})}\BibitemShut {NoStop}%
\bibitem [{\citenamefont {Reivelt}\ and\ \citenamefont
  {Saari}(2003)}]{Reivelt03arxiv}%
  \BibitemOpen
  \bibfield  {author} {\bibinfo {author} {\bibfnamefont {K.}~\bibnamefont
  {Reivelt}}\ and\ \bibinfo {author} {\bibfnamefont {P.}~\bibnamefont
  {Saari}},\ }\href@noop {} {\bibfield  {journal} {\bibinfo  {journal}
  {arxiv:physics/0309079}\ } (\bibinfo {year} {2003})}\BibitemShut {NoStop}%
\bibitem [{\citenamefont {Kiselev}(2007)}]{Kiselev07OS}%
  \BibitemOpen
  \bibfield  {author} {\bibinfo {author} {\bibfnamefont {A.~P.}\ \bibnamefont
  {Kiselev}},\ }\href@noop {} {\bibfield  {journal} {\bibinfo  {journal} {Opt.
  Spectrosc.}\ }\textbf {\bibinfo {volume} {102}},\ \bibinfo {pages} {603}
  (\bibinfo {year} {2007})}\BibitemShut {NoStop}%
\bibitem [{\citenamefont {Turunen}\ and\ \citenamefont
  {Friberg}(2010)}]{Turunen10PO}%
  \BibitemOpen
  \bibfield  {author} {\bibinfo {author} {\bibfnamefont {J.}~\bibnamefont
  {Turunen}}\ and\ \bibinfo {author} {\bibfnamefont {A.~T.}\ \bibnamefont
  {Friberg}},\ }\href@noop {} {\bibfield  {journal} {\bibinfo  {journal} {Prog.
  Opt.}\ }\textbf {\bibinfo {volume} {54}},\ \bibinfo {pages} {1} (\bibinfo
  {year} {2010})}\BibitemShut {NoStop}%
\bibitem [{\citenamefont {Hern\'andez-Figueroa}\ \emph
  {et~al.}(2014)\citenamefont {Hern\'andez-Figueroa}, \citenamefont {Recami},\
  and\ \citenamefont {Zamboni-Rached}}]{FigueroaBook14}%
  \BibitemOpen
  \bibinfo {editor} {\bibfnamefont {H.~E.}\ \bibnamefont
  {Hern\'andez-Figueroa}}, \bibinfo {editor} {\bibfnamefont {E.}~\bibnamefont
  {Recami}}, \ and\ \bibinfo {editor} {\bibfnamefont {M.}~\bibnamefont
  {Zamboni-Rached}},\ eds.,\ \href@noop {} {\emph {\bibinfo {title}
  {Non-diffracting Waves}}}\ (\bibinfo  {publisher} {Wiley-VCH},\ \bibinfo
  {year} {2014})\BibitemShut {NoStop}%
\bibitem [{\citenamefont {Saari}\ and\ \citenamefont
  {Reivelt}(1997)}]{Saari97PRL}%
  \BibitemOpen
  \bibfield  {author} {\bibinfo {author} {\bibfnamefont {P.}~\bibnamefont
  {Saari}}\ and\ \bibinfo {author} {\bibfnamefont {K.}~\bibnamefont
  {Reivelt}},\ }\href@noop {} {\bibfield  {journal} {\bibinfo  {journal} {Phys.
  Rev. Lett.}\ }\textbf {\bibinfo {volume} {79}},\ \bibinfo {pages} {4135}
  (\bibinfo {year} {1997})}\BibitemShut {NoStop}%
\bibitem [{\citenamefont {Reivelt}\ and\ \citenamefont
  {Saari}(2000)}]{Reivelt00JOSAA}%
  \BibitemOpen
  \bibfield  {author} {\bibinfo {author} {\bibfnamefont {K.}~\bibnamefont
  {Reivelt}}\ and\ \bibinfo {author} {\bibfnamefont {P.}~\bibnamefont
  {Saari}},\ }\href@noop {} {\bibfield  {journal} {\bibinfo  {journal} {J. Opt.
  Soc. Am. A}\ }\textbf {\bibinfo {volume} {17}},\ \bibinfo {pages} {1785}
  (\bibinfo {year} {2000})}\BibitemShut {NoStop}%
\bibitem [{\citenamefont {Reivelt}\ and\ \citenamefont
  {Saari}(2002)}]{Reivelt02PRE}%
  \BibitemOpen
  \bibfield  {author} {\bibinfo {author} {\bibfnamefont {K.}~\bibnamefont
  {Reivelt}}\ and\ \bibinfo {author} {\bibfnamefont {P.}~\bibnamefont
  {Saari}},\ }\href@noop {} {\bibfield  {journal} {\bibinfo  {journal} {Phys.
  Rev. E}\ }\textbf {\bibinfo {volume} {66}},\ \bibinfo {pages} {056611}
  (\bibinfo {year} {2002})}\BibitemShut {NoStop}%
\bibitem [{\citenamefont {Kondakci}\ and\ \citenamefont
  {Abouraddy}(2017)}]{Kondakci17NP}%
  \BibitemOpen
  \bibfield  {author} {\bibinfo {author} {\bibfnamefont {H.~E.}\ \bibnamefont
  {Kondakci}}\ and\ \bibinfo {author} {\bibfnamefont {A.~F.}\ \bibnamefont
  {Abouraddy}},\ }\href@noop {} {\bibfield  {journal} {\bibinfo  {journal}
  {Nat. Photon.}\ }\textbf {\bibinfo {volume} {11}},\ \bibinfo {pages} {733}
  (\bibinfo {year} {2017})}\BibitemShut {NoStop}%
\bibitem [{\citenamefont {Yessenov}\ \emph
  {et~al.}(2019{\natexlab{a}})\citenamefont {Yessenov}, \citenamefont
  {Bhaduri}, \citenamefont {Kondakci},\ and\ \citenamefont
  {Abouraddy}}]{Yessenov19OPN}%
  \BibitemOpen
  \bibfield  {author} {\bibinfo {author} {\bibfnamefont {M.}~\bibnamefont
  {Yessenov}}, \bibinfo {author} {\bibfnamefont {B.}~\bibnamefont {Bhaduri}},
  \bibinfo {author} {\bibfnamefont {H.~E.}\ \bibnamefont {Kondakci}}, \ and\
  \bibinfo {author} {\bibfnamefont {A.~F.}\ \bibnamefont {Abouraddy}},\
  }\href@noop {} {\bibfield  {journal} {\bibinfo  {journal} {Opt. Photon.
  News}\ }\textbf {\bibinfo {volume} {30}},\ \bibinfo {pages} {34} (\bibinfo
  {year} {2019}{\natexlab{a}})}\BibitemShut {NoStop}%
\bibitem [{\citenamefont {Kondakci}\ and\ \citenamefont
  {Abouraddy}(2019)}]{Kondakci19NC}%
  \BibitemOpen
  \bibfield  {author} {\bibinfo {author} {\bibfnamefont {H.~E.}\ \bibnamefont
  {Kondakci}}\ and\ \bibinfo {author} {\bibfnamefont {A.~F.}\ \bibnamefont
  {Abouraddy}},\ }\href@noop {} {\bibfield  {journal} {\bibinfo  {journal}
  {Nat. Commun.}\ }\textbf {\bibinfo {volume} {10}},\ \bibinfo {pages} {929}
  (\bibinfo {year} {2019})}\BibitemShut {NoStop}%
\bibitem [{\citenamefont {Bhaduri}\ \emph {et~al.}(2019)\citenamefont
  {Bhaduri}, \citenamefont {Yessenov},\ and\ \citenamefont
  {Abouraddy}}]{Bhaduri19Optica}%
  \BibitemOpen
  \bibfield  {author} {\bibinfo {author} {\bibfnamefont {B.}~\bibnamefont
  {Bhaduri}}, \bibinfo {author} {\bibfnamefont {M.}~\bibnamefont {Yessenov}}, \
  and\ \bibinfo {author} {\bibfnamefont {A.~F.}\ \bibnamefont {Abouraddy}},\
  }\href@noop {} {\bibfield  {journal} {\bibinfo  {journal} {Optica}\ }\textbf
  {\bibinfo {volume} {6}},\ \bibinfo {pages} {139} (\bibinfo {year}
  {2019})}\BibitemShut {NoStop}%
\bibitem [{\citenamefont {Kondakci}\ and\ \citenamefont
  {Abouraddy}(2018{\natexlab{a}})}]{Kondakci18OL}%
  \BibitemOpen
  \bibfield  {author} {\bibinfo {author} {\bibfnamefont {H.~E.}\ \bibnamefont
  {Kondakci}}\ and\ \bibinfo {author} {\bibfnamefont {A.~F.}\ \bibnamefont
  {Abouraddy}},\ }\href@noop {} {\bibfield  {journal} {\bibinfo  {journal}
  {Opt. Lett.}\ }\textbf {\bibinfo {volume} {43}},\ \bibinfo {pages} {3830}
  (\bibinfo {year} {2018}{\natexlab{a}})}\BibitemShut {NoStop}%
\bibitem [{\citenamefont {Kondakci}\ and\ \citenamefont
  {Abouraddy}(2018{\natexlab{b}})}]{Kondakci18PRL}%
  \BibitemOpen
  \bibfield  {author} {\bibinfo {author} {\bibfnamefont {H.~E.}\ \bibnamefont
  {Kondakci}}\ and\ \bibinfo {author} {\bibfnamefont {A.~F.}\ \bibnamefont
  {Abouraddy}},\ }\href@noop {} {\bibfield  {journal} {\bibinfo  {journal}
  {Phys. Rev. Lett.}\ }\textbf {\bibinfo {volume} {120}},\ \bibinfo {pages}
  {163901} (\bibinfo {year} {2018}{\natexlab{b}})}\BibitemShut {NoStop}%
\bibitem [{\citenamefont {Yessenov}\ \emph
  {et~al.}(2019{\natexlab{b}})\citenamefont {Yessenov}, \citenamefont
  {Bhaduri}, \citenamefont {Kondakci}, \citenamefont {Meem}, \citenamefont
  {Menon},\ and\ \citenamefont {Abouraddy}}]{Yessenov19Optica}%
  \BibitemOpen
  \bibfield  {author} {\bibinfo {author} {\bibfnamefont {M.}~\bibnamefont
  {Yessenov}}, \bibinfo {author} {\bibfnamefont {B.}~\bibnamefont {Bhaduri}},
  \bibinfo {author} {\bibfnamefont {H.~E.}\ \bibnamefont {Kondakci}}, \bibinfo
  {author} {\bibfnamefont {M.}~\bibnamefont {Meem}}, \bibinfo {author}
  {\bibfnamefont {R.}~\bibnamefont {Menon}}, \ and\ \bibinfo {author}
  {\bibfnamefont {A.~F.}\ \bibnamefont {Abouraddy}},\ }\href@noop {} {\bibfield
   {journal} {\bibinfo  {journal} {Optica}\ }\textbf {\bibinfo {volume} {6}},\
  \bibinfo {pages} {522} (\bibinfo {year} {2019}{\natexlab{b}})}\BibitemShut
  {NoStop}%
\bibitem [{\citenamefont {Yessenov}\ and\ \citenamefont
  {Abouraddy}(2019)}]{Yessenov19OL}%
  \BibitemOpen
  \bibfield  {author} {\bibinfo {author} {\bibfnamefont {M.}~\bibnamefont
  {Yessenov}}\ and\ \bibinfo {author} {\bibfnamefont {A.~F.}\ \bibnamefont
  {Abouraddy}},\ }\href@noop {} {\bibfield  {journal} {\bibinfo  {journal}
  {Opt. Lett.}\ }\textbf {\bibinfo {volume} {44}},\ \bibinfo {pages} {5125}
  (\bibinfo {year} {2019})}\BibitemShut {NoStop}%
\bibitem [{\citenamefont {Villinger}\ \emph {et~al.}(2019)\citenamefont
  {Villinger}, \citenamefont {Shiri}, \citenamefont {Shabahang}, \citenamefont
  {Jahromi}, \citenamefont {Nasr}, \citenamefont {Villinger},\ and\
  \citenamefont {Abouraddy}}]{Villinger19unpubl}%
  \BibitemOpen
  \bibfield  {author} {\bibinfo {author} {\bibfnamefont {M.~L.}\ \bibnamefont
  {Villinger}}, \bibinfo {author} {\bibfnamefont {A.}~\bibnamefont {Shiri}},
  \bibinfo {author} {\bibfnamefont {S.}~\bibnamefont {Shabahang}}, \bibinfo
  {author} {\bibfnamefont {A.~K.}\ \bibnamefont {Jahromi}}, \bibinfo {author}
  {\bibfnamefont {M.~B.}\ \bibnamefont {Nasr}}, \bibinfo {author}
  {\bibfnamefont {C.}~\bibnamefont {Villinger}}, \ and\ \bibinfo {author}
  {\bibfnamefont {A.~F.}\ \bibnamefont {Abouraddy}},\ }\href@noop {} {\bibfield
   {journal} {\bibinfo  {journal} {arXiv:1911.09276}\ } (\bibinfo {year}
  {2019})}\BibitemShut {NoStop}%
\bibitem [{\citenamefont {Wicht}\ \emph {et~al.}(1997)\citenamefont {Wicht},
  \citenamefont {Danzmann}, \citenamefont {Fleischhauer}, \citenamefont
  {Scully}, \citenamefont {M{\"u}ller},\ and\ \citenamefont
  {Rinkleff}}]{Wicht97OC}%
  \BibitemOpen
  \bibfield  {author} {\bibinfo {author} {\bibfnamefont {A.}~\bibnamefont
  {Wicht}}, \bibinfo {author} {\bibfnamefont {K.}~\bibnamefont {Danzmann}},
  \bibinfo {author} {\bibfnamefont {M.}~\bibnamefont {Fleischhauer}}, \bibinfo
  {author} {\bibfnamefont {M.}~\bibnamefont {Scully}}, \bibinfo {author}
  {\bibfnamefont {G.}~\bibnamefont {M{\"u}ller}}, \ and\ \bibinfo {author}
  {\bibfnamefont {R.-H.}\ \bibnamefont {Rinkleff}},\ }\href@noop {} {\bibfield
  {journal} {\bibinfo  {journal} {Opt. Commun.}\ }\textbf {\bibinfo {volume}
  {134}},\ \bibinfo {pages} {431} (\bibinfo {year} {1997})}\BibitemShut
  {NoStop}%
\bibitem [{\citenamefont {Pati}\ \emph {et~al.}(2007)\citenamefont {Pati},
  \citenamefont {Salit}, \citenamefont {Salit},\ and\ \citenamefont
  {Shahriar}}]{Pati07PRL}%
  \BibitemOpen
  \bibfield  {author} {\bibinfo {author} {\bibfnamefont {G.~S.}\ \bibnamefont
  {Pati}}, \bibinfo {author} {\bibfnamefont {M.}~\bibnamefont {Salit}},
  \bibinfo {author} {\bibfnamefont {K.}~\bibnamefont {Salit}}, \ and\ \bibinfo
  {author} {\bibfnamefont {M.~S.}\ \bibnamefont {Shahriar}},\ }\href@noop {}
  {\bibfield  {journal} {\bibinfo  {journal} {Phys. Rev. Lett.}\ }\textbf
  {\bibinfo {volume} {99}},\ \bibinfo {pages} {133601} (\bibinfo {year}
  {2007})}\BibitemShut {NoStop}%
\bibitem [{\citenamefont {Wu}\ and\ \citenamefont {Xiao}(2008)}]{Wu08PRA}%
  \BibitemOpen
  \bibfield  {author} {\bibinfo {author} {\bibfnamefont {H.}~\bibnamefont
  {Wu}}\ and\ \bibinfo {author} {\bibfnamefont {M.}~\bibnamefont {Xiao}},\
  }\href@noop {} {\bibfield  {journal} {\bibinfo  {journal} {Phys. Rev. A}\
  }\textbf {\bibinfo {volume} {77}},\ \bibinfo {pages} {031801(R)} (\bibinfo
  {year} {2008})}\BibitemShut {NoStop}%
\bibitem [{\citenamefont {Yum}\ \emph {et~al.}(2013{\natexlab{a}})\citenamefont
  {Yum}, \citenamefont {Sheuer}, \citenamefont {Salit}, \citenamefont
  {Hemmer},\ and\ \citenamefont {Shahriar}}]{Yum13JLT}%
  \BibitemOpen
  \bibfield  {author} {\bibinfo {author} {\bibfnamefont {H.~N.}\ \bibnamefont
  {Yum}}, \bibinfo {author} {\bibfnamefont {J.}~\bibnamefont {Sheuer}},
  \bibinfo {author} {\bibfnamefont {M.}~\bibnamefont {Salit}}, \bibinfo
  {author} {\bibfnamefont {P.~R.}\ \bibnamefont {Hemmer}}, \ and\ \bibinfo
  {author} {\bibfnamefont {M.~S.}\ \bibnamefont {Shahriar}},\ }\href@noop {}
  {\bibfield  {journal} {\bibinfo  {journal} {J. Lightwave Technol.}\ }\textbf
  {\bibinfo {volume} {32}},\ \bibinfo {pages} {3865} (\bibinfo {year}
  {2013}{\natexlab{a}})}\BibitemShut {NoStop}%
\bibitem [{\citenamefont {Wise}\ \emph {et~al.}(2005)\citenamefont {Wise},
  \citenamefont {Quetschke}, \citenamefont {Deshpande}, \citenamefont
  {Mueller}, \citenamefont {Reitze}, \citenamefont {Tanner}, \citenamefont
  {Whiting}, \citenamefont {Chen}, \citenamefont {T{\"u}nnermann},
  \citenamefont {Kley},\ and\ \citenamefont {Clausnitzer}}]{Wise05PRL}%
  \BibitemOpen
  \bibfield  {author} {\bibinfo {author} {\bibfnamefont {S.}~\bibnamefont
  {Wise}}, \bibinfo {author} {\bibfnamefont {V.}~\bibnamefont {Quetschke}},
  \bibinfo {author} {\bibfnamefont {A.~J.}\ \bibnamefont {Deshpande}}, \bibinfo
  {author} {\bibfnamefont {G.}~\bibnamefont {Mueller}}, \bibinfo {author}
  {\bibfnamefont {D.~H.}\ \bibnamefont {Reitze}}, \bibinfo {author}
  {\bibfnamefont {D.~B.}\ \bibnamefont {Tanner}}, \bibinfo {author}
  {\bibfnamefont {B.~F.}\ \bibnamefont {Whiting}}, \bibinfo {author}
  {\bibfnamefont {Y.}~\bibnamefont {Chen}}, \bibinfo {author} {\bibfnamefont
  {A.}~\bibnamefont {T{\"u}nnermann}}, \bibinfo {author} {\bibfnamefont
  {E.}~\bibnamefont {Kley}}, \ and\ \bibinfo {author} {\bibfnamefont
  {T.}~\bibnamefont {Clausnitzer}},\ }\href@noop {} {\bibfield  {journal}
  {\bibinfo  {journal} {Phys. Rev. Lett.}\ }\textbf {\bibinfo {volume} {95}},\
  \bibinfo {pages} {013901} (\bibinfo {year} {2005})}\BibitemShut {NoStop}%
\bibitem [{\citenamefont {Yum}\ \emph {et~al.}(2013{\natexlab{b}})\citenamefont
  {Yum}, \citenamefont {Liu}, \citenamefont {Hemmer}, \citenamefont {Scheuer},\
  and\ \citenamefont {Shahriar}}]{Yum13OC}%
  \BibitemOpen
  \bibfield  {author} {\bibinfo {author} {\bibfnamefont {H.~N.}\ \bibnamefont
  {Yum}}, \bibinfo {author} {\bibfnamefont {X.}~\bibnamefont {Liu}}, \bibinfo
  {author} {\bibfnamefont {P.~R.}\ \bibnamefont {Hemmer}}, \bibinfo {author}
  {\bibfnamefont {J.}~\bibnamefont {Scheuer}}, \ and\ \bibinfo {author}
  {\bibfnamefont {M.~S.}\ \bibnamefont {Shahriar}},\ }\href@noop {} {\bibfield
  {journal} {\bibinfo  {journal} {Opt. Commun.}\ }\textbf {\bibinfo {volume}
  {305}},\ \bibinfo {pages} {260} (\bibinfo {year}
  {2013}{\natexlab{b}})}\BibitemShut {NoStop}%
\bibitem [{\citenamefont {Shabahang}\ \emph {et~al.}(2017)\citenamefont
  {Shabahang}, \citenamefont {Kondakci}, \citenamefont {Villinger},
  \citenamefont {Perlstein}, \citenamefont {{El H}alawany},\ and\ \citenamefont
  {Abouraddy}}]{Shabahang17SR}%
  \BibitemOpen
  \bibfield  {author} {\bibinfo {author} {\bibfnamefont {S.}~\bibnamefont
  {Shabahang}}, \bibinfo {author} {\bibfnamefont {H.~E.}\ \bibnamefont
  {Kondakci}}, \bibinfo {author} {\bibfnamefont {M.~L.}\ \bibnamefont
  {Villinger}}, \bibinfo {author} {\bibfnamefont {J.~D.}\ \bibnamefont
  {Perlstein}}, \bibinfo {author} {\bibfnamefont {A.}~\bibnamefont {{El
  H}alawany}}, \ and\ \bibinfo {author} {\bibfnamefont {A.~F.}\ \bibnamefont
  {Abouraddy}},\ }\href@noop {} {\bibfield  {journal} {\bibinfo  {journal}
  {Sci. Rep.}\ }\textbf {\bibinfo {volume} {7}},\ \bibinfo {pages} {10336}
  (\bibinfo {year} {2017})}\BibitemShut {NoStop}%
\bibitem [{\citenamefont {Shabahang}\ \emph {et~al.}(2019)\citenamefont
  {Shabahang}, \citenamefont {Jahromi}, \citenamefont {Shiri}, \citenamefont
  {Schepler},\ and\ \citenamefont {Abouraddy}}]{Shabahang19OL}%
  \BibitemOpen
  \bibfield  {author} {\bibinfo {author} {\bibfnamefont {S.}~\bibnamefont
  {Shabahang}}, \bibinfo {author} {\bibfnamefont {A.~K.}\ \bibnamefont
  {Jahromi}}, \bibinfo {author} {\bibfnamefont {A.}~\bibnamefont {Shiri}},
  \bibinfo {author} {\bibfnamefont {K.~L.}\ \bibnamefont {Schepler}}, \ and\
  \bibinfo {author} {\bibfnamefont {A.~F.}\ \bibnamefont {Abouraddy}},\
  }\href@noop {} {\bibfield  {journal} {\bibinfo  {journal} {Opt. Lett.}\
  }\textbf {\bibinfo {volume} {44}},\ \bibinfo {pages} {1532} (\bibinfo {year}
  {2019})}\BibitemShut {NoStop}%
\bibitem [{\citenamefont {Zapata-Rodr{\'i}guez}\ and\ \citenamefont
  {Porras}(2006)}]{Zapata06OL}%
  \BibitemOpen
  \bibfield  {author} {\bibinfo {author} {\bibfnamefont {C.~J.}\ \bibnamefont
  {Zapata-Rodr{\'i}guez}}\ and\ \bibinfo {author} {\bibfnamefont {M.~A.}\
  \bibnamefont {Porras}},\ }\href@noop {} {\bibfield  {journal} {\bibinfo
  {journal} {Opt. Lett.}\ }\textbf {\bibinfo {volume} {31}},\ \bibinfo {pages}
  {3532} (\bibinfo {year} {2006})}\BibitemShut {NoStop}%
\bibitem [{\citenamefont {Shaarawi}\ and\ \citenamefont
  {Besieris}(2000)}]{Shaarawi00JPA}%
  \BibitemOpen
  \bibfield  {author} {\bibinfo {author} {\bibfnamefont {A.~M.}\ \bibnamefont
  {Shaarawi}}\ and\ \bibinfo {author} {\bibfnamefont {I.~M.}\ \bibnamefont
  {Besieris}},\ }\href@noop {} {\bibfield  {journal} {\bibinfo  {journal} {J.
  Phys. A}\ }\textbf {\bibinfo {volume} {33}},\ \bibinfo {pages} {7255}
  (\bibinfo {year} {2000})}\BibitemShut {NoStop}%
\bibitem [{\citenamefont {Saari}(2018)}]{SaariPRA18}%
  \BibitemOpen
  \bibfield  {author} {\bibinfo {author} {\bibfnamefont {P.}~\bibnamefont
  {Saari}},\ }\href@noop {} {\bibfield  {journal} {\bibinfo  {journal} {Phys.
  Rev. A}\ }\textbf {\bibinfo {volume} {97}},\ \bibinfo {pages} {063824}
  (\bibinfo {year} {2018})}\BibitemShut {NoStop}%
\bibitem [{\citenamefont {Yessenov}\ \emph
  {et~al.}(2019{\natexlab{c}})\citenamefont {Yessenov}, \citenamefont
  {Bhaduri}, \citenamefont {Mach}, \citenamefont {Mardani}, \citenamefont
  {Kondakci}, \citenamefont {Alonso}, \citenamefont {Atia},\ and\ \citenamefont
  {Abouraddy}}]{Yessenov19OE}%
  \BibitemOpen
  \bibfield  {author} {\bibinfo {author} {\bibfnamefont {M.}~\bibnamefont
  {Yessenov}}, \bibinfo {author} {\bibfnamefont {B.}~\bibnamefont {Bhaduri}},
  \bibinfo {author} {\bibfnamefont {L.}~\bibnamefont {Mach}}, \bibinfo {author}
  {\bibfnamefont {D.}~\bibnamefont {Mardani}}, \bibinfo {author} {\bibfnamefont
  {H.~E.}\ \bibnamefont {Kondakci}}, \bibinfo {author} {\bibfnamefont {M.~A.}\
  \bibnamefont {Alonso}}, \bibinfo {author} {\bibfnamefont {G.~A.}\
  \bibnamefont {Atia}}, \ and\ \bibinfo {author} {\bibfnamefont {A.~F.}\
  \bibnamefont {Abouraddy}},\ }\href@noop {} {\bibfield  {journal} {\bibinfo
  {journal} {Opt. Express}\ }\textbf {\bibinfo {volume} {27}},\ \bibinfo
  {pages} {12443} (\bibinfo {year} {2019}{\natexlab{c}})}\BibitemShut {NoStop}%
\bibitem [{\citenamefont {Yessenov}\ \emph
  {et~al.}(2019{\natexlab{d}})\citenamefont {Yessenov}, \citenamefont
  {Bhaduri}, \citenamefont {Kondakci},\ and\ \citenamefont
  {Abouraddy}}]{Yessenov19PRA}%
  \BibitemOpen
  \bibfield  {author} {\bibinfo {author} {\bibfnamefont {M.}~\bibnamefont
  {Yessenov}}, \bibinfo {author} {\bibfnamefont {B.}~\bibnamefont {Bhaduri}},
  \bibinfo {author} {\bibfnamefont {H.~E.}\ \bibnamefont {Kondakci}}, \ and\
  \bibinfo {author} {\bibfnamefont {A.~F.}\ \bibnamefont {Abouraddy}},\
  }\href@noop {} {\bibfield  {journal} {\bibinfo  {journal} {Phys. Rev. A}\
  }\textbf {\bibinfo {volume} {99}},\ \bibinfo {pages} {023856} (\bibinfo
  {year} {2019}{\natexlab{d}})}\BibitemShut {NoStop}%
\end{thebibliography}%

%\bibliographyfullrefs{diffraction}

\end{document}